# Control of LED Emission with Functional Dielectric Metasurfaces

*Egor Khaidarov[†], Zhengtong Liu[†*], Ramón Paniagua-Domínguez, Son Tung Ha, Vytautas Valuckas, Xinan Liang, Yuriy Akimov, Ping Bai, Ching Eng Png, Hilmi Volkan Demir and Arseniy I. Kuznetsov[*]*

E. Khaidarov, Dr. R. Paniagua-Domínguez, Dr. S. T. Ha, V. Valuckas, Dr. X. Liang, Dr. A. I. Kuznetsov*

Institute of Materials Research and Engineering (Agency for Science, Technology and Research, A*STAR)

2 Fusionopolis Way, #08-03, Innovis 138634, Singapore

E-mail: arseniy_kuznetsov@imre.a-star.edu.sg

E. Khaidarov, Prof. H. V. Demir

LUMINOUS! Center of Excellence for Semiconductor Lighting and Displays, The Photonics Institute, School of Electrical and Electronic Engineering, School of Physical and Mathematical Sciences, Nanyang Technological University

50 Nanyang Avenue, 639798, Singapore

Dr. Z. Liu*, Dr. Y. Akimov, Dr. P. Bai, Dr. C. E. Png

Institute of High Performance Computing, A*STAR (Agency for Science, Technology and Research)

1 Fusionopolis Way, #16-16 Connexis North, 138632, Singapore

E-mail: liuz@ihpc.a-star.edu.sg

Prof. H. V. Demir

Bilkent University UNAM, Institute of Materials Science and Nanotechnology

Ankara, 06800, Turkey

[†]These authors contributed equally to this work.

Abstract

The improvement of light-emitting diodes (LEDs) is one of the major goals of optoelectronics and photonics research. While emission rate enhancement is certainly one of the targets, in this regard, for LED integration to complex photonic devices, one would require to have, additionally, precise control of the wavefront of the emitted light. Metasurfaces are spatial arrangements of engineered scatters that may enable this light manipulation capability with unprecedented resolution. Most of these devices, however, are only able to function properly under irradiation of light with a large spatial coherence, typically normally incident lasers. LEDs, on the other hand, have angularly broad, Lambertian-like emission patterns characterized by a low spatial coherence, which makes the integration of metasurface devices on LED architectures extremely challenging. A novel concept for metasurface integration on LED is proposed, using a cavity to increase the LED spatial coherence through an angular collimation. Due to the resonant character of the cavity, extending the spatial coherence of the emitted light does not come at the price of any reduction in the total emitted power. The experimental demonstration of the proposed concept is implemented on a GaP LED architecture including a hybrid metallic-Bragg cavity. By integrating a silicon metasurface on top we demonstrate two different functionalities of these compact devices: directional LED emission at a desired angle and LED emission of a vortex beam with an orbital angular momentum. The presented concept is general, being applicable to other incoherent light sources and enabling metasurfaces designed for plane waves to work with incoherent light emitters.

Since their discovery in the early years of the 20th century, light-emitting diodes (LEDs) have found numerous applications as efficient, low-cost and compact light sources.[1,2] Due to their wide use in large business areas, such as lighting and displays, as well as their potential application in novel directions such as Li-Fi and photonic interconnects, they still attract strong interest of the research community. Although the various types of inorganic, organic and quantum-dot-based LEDs have their own peculiarities, advantages and disadvantages, most of the research towards their improvement shares similar goals. In particular, increasing the speed of these devices by reducing the recombination lifetimes via radiative channels and enhancing their efficiency are among the top priorities in the field. In particular, regarding the efficiency, the main efforts focus towards improving the external quantum efficiency (EQE), defined as the number of photons emitted from the device per number of electrons injected into the device and largely determined by device design and architecture through the extraction efficiency. For a planar interface, the amount of photons that can escape with respect to the total amount generated is limited by their total internal reflection at the interface between the LED and the external medium and, therefore, depends solely on the refractive index contrast between these two. With this is mind, a simple way to improve the EQE is patterning the surface of the device with different scatterers.[3,4] These scatterers can be precisely engineered, e.g. in the form of photonic crystal structures, to extract part of the energy that otherwise would be simply coupled into slab modes – and hence lost – and convert it to propagating light in the external medium, thus, adding to the total efficiency of the device.[5] Even more, when properly designed, these engineered structures[6] (as well as others such as plasmonic particles[7,8] or grooves[9]) may also change the emission pattern of the LED device (or other random light sources[10,11]), an effect that can be used to produce directional light sources that are, as well, compact, avoiding the use of external elements such as mirrors or lenses. This directional emission endows LEDs with an attractive, yet simple, functionality for integrated devices.

Looking forward, achieving directional emission from the LED is insufficient for more complex applications and further device integration. These, in general, require precise control of the wavefront of emitted light. Recently, a mean to achieve this degree of control has emerged by introducing abrupt discontinuities in the phase of the incident wave using arrangements of engineered scatterers[12,13], so-called metasurfaces. Among their different possible realizations, dielectric metasurfaces[14,15], with low losses, have emerged as a one of the best solutions to obtain high efficiency devices. On top of that, the rich phenomenology of optical modes excited in dielectric particles[14] has enabled new effects and applications for metasurfaces based on them. Novel effects include, e.g., the realization of optical Huygens' metasurfaces,[16,17,18] generalized version of the Brewster effect,[19,20] nonradiating anapole modes[21,22] and quasi-bound states in the continuum[23-25] The range of applications, in turn, includes achieving high angle light beam bending devices.[26-28] and polarization beam splitters[29,30], obtaining ultra-high numerical aperture,[31] tunable,[32] low reflection and/or broadband lenses,[33] generation of complex holograms[34,35] and vortex beams,[36] as well as the realization of near-zero refractive index materials[37,38] and compact cameras, microscopy[39] and spectrometers,[40,41] to mention some.

In most of these applications, however, the designed metasurfaces require a spatially coherent illumination, most of them working under plane wave excitation (with very few exception in which metasurfaces are designed for point-like sources with well-defined positions[42,43]). On the contrary, a typical LED exhibits Lambertian-shaped emission profile with very low spatial coherence. This low coherence is particularly important if the metasurface and the LED source have comparable sizes and are placed in close proximity (several wavelengths) of each other - a situation desirable for compact integration. Consequently, introduction of structural modifications becomes necessary to obtain a combined functional device. This is the reason why, so far, successful integration of metasurface devices on the LED has been limited to control of the directionality and/or polarization[9,44,45] of emitted light, while full wavefront control remains elusive.

In this work, we propose a new method for direct integration of metasurface devices on top of an LED. The method relies on a collimating element, in particular, a resonant cavity. Both the metasurface and the cavity LED can be designed separately and, thus, the method allows integration of any functional metasurface designed to work under normally incident plane wave illumination, provided that it works in the transmission regime. Moreover, due to its resonant character, the presence of the cavity enhances the emission of the device. We demonstrate experimentally the proposed concept realizing an efficient beam deflection device with a pre-designed emission angle, which closely matches the theoretical predictions. Furthermore, to demonstrate the full power of this approach, which enables total wavefront control, we demonstrate generation of LED light with orbital angular momentum by integrating a metasurface designed to generate a vortex beam.

The proposed method is schematically depicted in **Figure 1a**. First, spatially incoherent emission of the LED is modified by restricting the initial broad Lambertian angular distribution to a narrow range of angles. This is achieved by introducing a resonant Fabry-Perot (FP) cavity into the LED architecture. Note that, the spatial coherence increased by resonant cavities[46] (which have been used in the LED community for more than two decades to enhance LED efficiencies[47-49]) could also be realized using other approaches, such as zero index materials[37], plasmonic[50] or photonic crystals.[6] In this work, we use the resonant FP cavity as it provides sufficient collimation, emission enhancement and does not require complex fabrication steps. Finally, the metasurface performing the desired functionality is integrated at the output of the cavity to realize the compact device.

For the theoretical analysis of the integrated device, we performed full numerical simulations based on the reciprocity principle,[51] as described in Experimental Section. This approach allows precise computation of the radiation pattern, the relative contribution of each polarization and the enhancement/suppression of the total emission relative to a plain LED. We choose to work with a GaP LED structure with an active layer consisting of multiple quantum wells (MQW) emitting at around 620 nm wavelength. Following the characteristics of the samples used in the subsequent experiment, we assume that the active layer thickness is 100 nm with its first interface located 200 nm below the top GaP surface. The radiation patterns for various LED configurations are given in Figures 1 b-d. The respective insets show the schematic side views of the considered structures. First, the Lambertian emission from a semi-infinite LED stack is calculated and plotted in Figure 1b. The red and green lines represent, respectively, the emission patterns of p-polarized and s-polarized light. To illustrate that direct integration of a metasurface (as designed for plane wave excitation) on top of the LED is not a feasible solution to preserve the metasurface functionality, we repeat the emission pattern calculation when a metasurface designed to deflect a normally incident plane wave is placed on top of the GaP LED. The metasurface consists of a rectangular lattice of periodically repeated supercells with 300 nm (sub-diffractive) and 1200 nm (diffractive) periods. The elements constituting the supercell are amorphous silicon cylinders with a height of 350 nm and a circular cross-section to guarantee polarization-independent operation. The diameter of the cylinders increases gradually within the supercell (114 nm, 124 nm, 133 nm and 170 nm), leading to an increasing phase accumulation under coherent plane wave illumination and mapping a tilted wavefront of a deflected beam. When illuminated by a normally incident plane wave at a wavelength of 620 nm, this metasurface is able to deflect it at an angle of 30° with a 94% relative efficiency and around 50% overall transmission for two orthogonal polarizations (see supporting information **Figure S1a**). However, when this metasurface is directly integrated on top of the GaP LED, simulation results show that the Lambertian emission pattern of the bare LED remains almost unaltered (plotted in Supplementary Figure S1b). This means that under the angularly broad illumination of an LED the metasurface loses its functionality.

In order to transform the Lambertian LED emission into the one having a narrower angular distribution, which can be manipulated by the metasurface, we use a hybrid Bragg-gold resonant FP cavity as a collimating element (see inset in Figure 1c). Such an asymmetric choice of the FP cavity mirror is dictated, on one hand, by the developed GaP wafer transfer to gold (which acts as the bottom mirror) and, on the other one, by the low losses of the Bragg mirror (acting as the top reflector). The gold layer is chosen to be several hundred nanometers thick to ensure a high reflection and the Bragg reflector consists of alternating $TiO_2$ and $SiO_2$ layers, chosen due to their high refractive index contrast and negligible losses in the visible spectral range. The thicknesses of the layers satisfy the quarter wavelength condition - 63 nm for $TiO_2$ and 107 nm for $SiO_2$. An analysis of the FP cavity performance as a function of the number of layers in the Bragg stack reveals that 2 pairs of layers are enough to ensure a sufficient emission collimation. The thickness of the GaP is then optimized to be 415 nm to support a Fabry-Perot resonance at the central wavelength of the LED emission (around 620 nm). The MQW layer thickness and location remains the same as before. The resulting emission diagram of this system, generically known as the resonant cavity LED (RCLED),[46-49] is shown in Figure 1c. As expected, in this situation the light is concentrated in a narrow angular range close to the surface normal, corresponding to those angles for which the cavity resonance condition is met.

As a last step, the beam deflecting silicon metasurface described above is now placed on top of the RCLED. A thin $SiO_2$ spacer (210 nm) is added between them to decouple the metasurface from the near-fields of the cavity and to serve as a low index substrate for the metasurface. The emission pattern from this device, depicted in Fig. 1d, shows a single, angularly narrow emission lobe with a clear maximum at around 30° with respect to the normal, thus confirming the validity of the proposed concept. Since we used cylinders with a circular cross section as building blocks of the metasurface, the beam deflection effect is almost polarization insensitive, as seen by comparing the red and green curves representing, respectively, the emission patterns of p-polarized and s-polarized light. An important aspect to note is that independent design of the cavity and the metasurface can only be applied to highly transmissive metasurfaces (or at least to those with low reflection). If that is not the case, the reflected light from the metasurface can affect the cavity performance. In that situation, a more precise engineering considering both the metasurface and the cavity as a whole is required.

For the sake of completeness, we show in supporting information **Figures S2 a,b** the performance of a device with a simpler cavity, consisting of two gold mirrors, which could be advantageous for electrical pumping. As can be seen, the performance of this device is very similar to the hybrid cavity case, but with a lower emission power. We finalize the theoretical characterization of the presented devices by calculating the total emitted power relative to the bare LED stack. This is done by integrating the directional emission over the top hemisphere. These results are presented in Figure S2c. The RCLED including the Bragg reflector gives an enhancement of the total emission of almost three times compared to the bare LED which is almost one order of magnitude higher compared to the RCLED based on gold mirrors. Including the metasurface on top of the RCLED slightly reduces, in both cases, the total emitted power. We attribute this to the non-negligible absorption of amorphous Si at the operating wavelength as well as to the small reflection from the metasurface (which was designed independently of the cavity). Nevertheless, the results still show that the total emitted power of the hybrid RCLED with the metasurface is almost twice higher than that of the bare LED and one order of magnitude higher than that of the configuration based on the gold FP cavity. Overall, both configurations - with gold or Bragg top reflector in the FP cavity - have their own advantages in terms of efficiency, design, fabrication simplicity and suitability for electrical pumping and, therefore, we believe that both architectures can find applications depending on the device requirements.

For the experimental demonstration of the proposed concept, we focus on the hybrid Bragg-gold cavity configuration. To realize the device, we start with a commercially available GaP p-i-n junction with multiple quantum wells with a total thickness of approximately 430 nm transferred to a gold-coated silicon substrate. The sample exhibits a photoluminescence band centered at 628 nm with a full-width at half-maximum (FWHM) of around 20 nm. To achieve the precise thickness of GaP, corresponding to the Fabry-Perot resonance, we use dry reactive ion etching (see Experimental Section for details). Next, two pairs of $TiO_2/SiO_2$ layers (63 nm /107 nm) are deposited on top using ion-assisted deposition, followed by a $SiO_2$ spacer layer (210 nm). The Si gradient metasurface is then fabricated on top of the device using electron beam lithography, as explained in details in Experimental Section. As described in the previous section, the metasurface supercell consists of four cylindrical elements and is designed to deflect the incoming wavefront at an angle of 30° with respect to the normal. The scanning electron microscopy (SEM) image of the metasurface fabricated on top of the RCLED is given in Figure 2c.

The optical characterization of the fabricated structures is performed using a spectrally-resolved back focal plane imaging setup,[24,28,30] as described in Experimental Section, incorporating a Nikon 100× microscope objective with a numerical aperture NA = 0.95. **Figure 2** presents a comparison of the angular-resolved reflection and photoluminescence spectra for the initial, uncovered LED slab on gold, the cavity LED incorporating the top Bragg mirror and the final device, with the integrated metasurface on top. The angular-resolved reflection spectra are obtained using the white light excitation. For the uncovered sample (Figure 2a) the measurements reveal two bands of slightly reduced reflectivity, corresponding to weak Fabry-Perot resonances that arise from the reflections between the gold substrate and the GaP-air interface. With the Bragg mirror on top, the resonant cavity LED exhibits much better pronounced Fabry-Perot resonances with almost vanishing reflectivity, one being centered at the emission wavelength of the GaP quantum wells of 628 nm (Figure 2b).

After characterizing the reflectivity of the device to corroborate the presence of the Fabry-Perot resonance at the desired working wavelength, we measure the photoluminescence (PL) signal by optically pumping the system with a continuous wave (CW) laser at 488 nm wavelength. We pump from the GaP side for the uncovered sample and from the Bragg reflector side for the RCLED. Figures 2d-f show the measured angularly-resolved PL spectra for all three LED architectures, namely, the uncovered LED and the RCLED without and with the metasurface on top. The angle represents the polar angle measured with respect to the normal direction to the sample surface. For the uncovered LED case, the PL emission is spread into all directions, including the large angles (Figure 2d), resembling the Lambertian pattern. A slight deviation is observed, however, as it is affected by the weak Fabry-Perot resonance observed in the reflection spectra, producing a slight narrowing of the angular emission distribution. On the other hand, the RCLED demonstrates a clearly enhanced directivity and angular narrowing (Figure 2e) as a result of the precise overlap of the PL emission maximum with the strong Fabry-Perot resonance.

When the beam deflection metasurface is incorporated on top of the RCLED, a very clear shift of the emission maximum towards the designed 30° angle is observed (Figure 2f). Note that, while in the first two cases the emission is azimuthally symmetric, this is not the case when the metasurface is incorporated on top. In this case, the PL shown in the figure corresponds to a cut along the long axis of the metasurface supercell. Figures 2g-i show the full angular distribution of unpolarized PL including all emission wavelengths. Note that panels 2d-f correspond to spectrally resolved cuts along the *y*-axis in panels 2g-i. Figure 2g indeed corroborates the quasi-Lambertian emission of the uncovered sample, while the hybrid Bragg-gold RCLED (Figure 2h) presents a clear directionality towards the normal to the sample surface. Finally, when the metasurface is integrated on top of the RCLED (Figure 2i), channeling of the LED emission towards the 30° angle is observed, as predicted by the metasurface design. It should be noted that, while numerical simulations of directivity were performed for a single wavelength, the experimental emission pattern contains a certain bandwidth. However, as it is a relatively narrow band (~20 nm), the device preserves the collimation and beam deflection functionality for the whole PL signal, including all wavelengths in the LED emission spectrum (Figure 2).

Nevertheless, a fairer comparison between the theory and experiment can be obtained by comparing the directionality cross sections at a specific wavelength (we chose, in particular, 634 nm wavelength providing the highest collimation). This comparison is shown in **Figure 3**. As can be seen from the figure, the experimental results have a very good correspondence with the theory, the main deviation being between the uncovered case and the pure Lambertian emission from a semi-infinite LED, stemming from the weak Fabry-Perot resonance observed in this system, as discussed above.

We finalize the characterization of the fabricated devices by estimating their relative emission efficiencies. This can be done by integrating the emission power in Figures 3g-i over the full measured solid angle. The directly obtained efficiencies are 39% for RCLED and 3% for the RCLED with the integrated metasurface, both relative to the plain GaP LED on gold. One should note that the emitted power is affected by the different pumping conditions for each device. While all of them are pumped from the same side, the pump beam experiences different reflections depending on whether it is the uncovered case, the case containing the Bragg stack, or that having both the stack and the metasurface. Thus, these directly obtained efficiencies should be normalized to the actual pumping power reaching the quantum wells in each case. Numerical simulations estimate that the real excitation power, relative to the uncovered case, is around 58% for the RCLED and 38% for RCLED with the integrated metasurface. In addition, simulations estimate an excitation enhancement of around 2.5 times for the uncovered case of the LED layer on gold compared to the bare semi-infinite LED. Taking into account these numbers we arrive at an estimated total emission efficiency of around 168% for the hybrid RCLED relative to the bare LED (without the gold substrate), and of around 20% when the metasurface is integrated on top. These values are still lower than those predicted by the simulations (shown in Figure S2c). This can be attributed to the wavelength dependence of the efficiency within the bandwidth of the Fabry-Perot resonance (~12 nm) which, moreover, is also narrower than the PL bandwidth (~20 nm), an effect not taken into account in the simulations. Underestimation of material losses in Si, Au, and the remaining EBL resist on top of the structures, as well as fabrication imperfections, might also affect the efficiency of the fabricated devices.

As a last step, we illustrate how the proposed concept allows to realize full wavefront control, and thus possible functionalities, which are not limited to the beam deflection. To do so, we fabricated and integrated on top of the same hybrid gold-Bragg RCLED another metasurface designed to generate a beam with an orbital angular momentum and topological charge of +1. Note that this functionality cannot be achieved unless full control of the wavefront is realized. The design of this metasurface follows the same approach of phase mapping, gradually varying the radius of the cylinders to achieve 0-2π phase variation along a full round of the azimuthal angle, thus imprinting the desired topological charge. The SEM image of this vortex beam generating metasurface fabricated on top of the RCLED is shown in supporting information **Figure S3a**. After pumping the device, we measure the intensity profile of the emitted light. The result is plotted in Figure S3b. As can be seen, the measured photoluminescence exhibits the characteristic "doughnut-shape" profile corresponding to a vortex beam with an orbital angular momentum and the expected topological charge, thus confirming the predicted functionality of the device.

In summary, we proposed a solution for integration of metasurfaces on top of LED to endow these devices with on-demand functionalities. The proposed solution is based on a collimating element, which can be realized, e.g., using a resonant cavity. This resonant cavity transforms the Lambertian-like emission from the bare LED into a directional one, thus having a large spatial coherence that can then be manipulated using any transmissive metasurface designed for a plane wave excitation. Both the cavity and the metasurface can, thus, be designed separately, as long as the reflection of the metasurface is low. Otherwise, the concept can still be used, provided that the system is optimized as a whole. Due to the resonant character of the cavity, the angular narrowing of the emission pattern does not come at the expense of a weaker emission, which can even be enhanced.

We experimentally demonstrated the proposed concept using a GaP-based LED combined with a hybrid gold-Bragg cavity and a silicon metasurface. We show two different functionalities, namely, a beam deflection and a vortex beam generation. However, the suggested method is general and can be applied to realize other, more complex functionalities such as LED beam focusing or hologram generation. Although demonstrated here using optical pumping, the concept can be readily adapted to obtain electrically pumped devices, either employing a fully metallic RCLED architecture, such as the one proposed based on gold layers, which corresponds to the flip-chip LED configuration, or substituting the bottom dielectric layer of the Bragg stack by some transparent conducting layer (e.g. ITO). Finally, we note that this concept can be applied to other random emitting media such as fluorescent molecules and quantum dots.

The proposed solution represents a step forward towards multipurpose, efficient, integrated optical devices, opening venues for compact integration of metasurfaces, which enable unprecedented control of light, with random sources. This can enable integration of LED sources instead of lasers into optoelectronic devices finding broad applications in optical communications, Li-Fi, displays, solid-state lighting, and many more.

**Experimental Section**

*Numerical simulations:* The Si metasurfaces on glass substrate were numerically simulated using the phase mapping approach with a plane wave excitation in commercially available Lumerical FDTD software. The unit cell boundary conditions for phase calculation were set to periodic and PMLs (Perfectly Matched Layers). The material data for the simulations was taken from the experimental ellipsometric measurements. The metasurfaces were designed separately from the LED.

The radiation pattern and extracted power were calculated using the reciprocity principle[51] using the commercially available COMSOL Multiphysics software. The proposed LED designs were inversely excited with a plane wave at different device emission angles. For each angle such backward propagating plane waves form a spatial power density distribution within the LED. The integral of this power density over the active region of the LED (MQW) is proportional to the direct power emission of the LED quantum wells at that specific angle.

*FP cavity fabrication:* GaP LEDs on gold substrates were purchased from Yangzhou Changelight Co. The stack of multiple quantum wells of around 100 nm in thickness is sandwiched in the GaP p-n junction. The whole LED stack thickness is around 430 nm. The thickness of the gold substrate is around 1 μm. The quantum wells emission is centered at 628 nm full-width-at-half-maximum (FWHM) of around 20 nm.

Although the thickness of the GaP layers can be precisely controlled during the growth, in case of a very thin LED layer the lattice mismatch with the substrate material deteriorates the quality of the film and affects the efficiency. This issue was solved by growing thicker structures and subsequently thinning them down by inductively coupled plasma-reactive ion etching (ICP-RIE, Oxford Plasmalab 100). We did that to obtain, with a high precision, the desired slab thickness corresponding to the low order Fabry-Perot resonance at the emission wavelength. We combined the dry etch based on Ar and Cl chemistry with reflection spectra measurements to adjust the thickness to approximately 416 nm. In those cases in which the GaP layer was slightly over-etched, a thin layer of amorphous Si was deposited to precisely meet the Fabry-Perot condition. For that we used a plasma enhanced chemical vapor deposition (PECVD, Oxford Plasmalab 100). The Bragg stack consisting of two alternating $TiO_2/SiO_2$ layer pairs (63 nm /107 nm each) was deposited using ion-assisted deposition (IAD Oxford Optofab 3000) with $SiO_2$ and $TiO_2$ targets. The 210 nm $SiO_2$ spacer layer was also deposited on top of the Bragg stack using IAD.

*Metasurface fabrication* For the metasurface fabrication we started with a deposition of an amorphous silicon film using the PECVD (Oxford Plasmalab 100) process. We subsequently spin-coated hydrogen silsesquioxane (HSQ, Dow Corning XR-1541) as a resist for the following electron beam lithography (EBL) process as well as a layer of Espacer 300AX01 (Showa Denko) as a charge dissipation layer. We then proceeded to expose the sample using an electron beam lithography system (EBL, Elionix ELS-7000, 100kV) to define the metasurface structure. Next, the sample was developed using a 25% solution of tetramethylammonium hydroxide (TMAH). Finally, the sample was etched using the same ICP-RIE tool with $Cl_2$ gas. The remaining HSQ resist on top of the etched structures was not removed, as the hydrofluoric acid (HF) solution commonly used for its removal might damage the $SiO_2$ substrate and the FP cavity stack.

*Back focal plane spectroscopy*

The angle-resolved reflection and photoluminescence measurements were performed in an inverted optical microscopy setup (Nikon Ti-U). A high power halogen light source (for reflection measurement) or continuous wave (CW) 488 nm laser (for photoluminescence measurement) irradiates the sample surface through the high NA objective (i.e., Nikon 100×,

0.95 NA). The reflection or emission signal is then collected by the same objective and routed to a high sensitivity, thermoelectrically cooled, charge coupled detector (Andor Newton EMCCD). Instead of an imaging plane, the back focal plane of the sample surface is projected onto a slit entrance (100 μm width) of the spectrometer (Andor SR-303i) with a 150 gr/mm grating (800 nm blading). This measurement technique gives both angular and spectral information along the slit axis of the response signal from the sample. For back focal plane image (without spectral information) the slit width is set at 2500 μm and the grating is set at the zero order (no diffraction). The maximum detection angle is determined by the NA of the objective, in this case approximately 72°.

**Supporting Information**

Supporting Information is available from the Wiley Online Library or from the author.

**Acknowledgments**

The authors acknowledge support from A*STAR SERC Pharos program, Grant No. 152 73 00025 (Singapore). E.K. acknowledges support from A*STAR Graduate Academy for the SINGA scholarship. H.V.D. acknowledges support from TUBA.

E. K., Z.L. contributed equally to this work.

**Author contributions**

Z.L., Y.A., H.V.D. and A. I. K. conceived the idea; E. K., Z.L. and Y.A. designed the metasurface and the FP cavity; Z.L., Y.A., E. K. and R. P.-D. performed the numerical simulations; E.K. fabricated the FP cavity and the metasurfaces on top of the LED; S.T.H. performed the optical measurements; X.L. measured the vortex beam emission; V. V. performed the SEM imaging; E. K. and Z.L. wrote the first draft of the manuscript. A. I. K., H.V.D., C.E.P, P.B. coordinated the work. All authors discussed the results and reviewed the manuscript.

**Conflict of Interest**

The authors declare no conflict of interest.

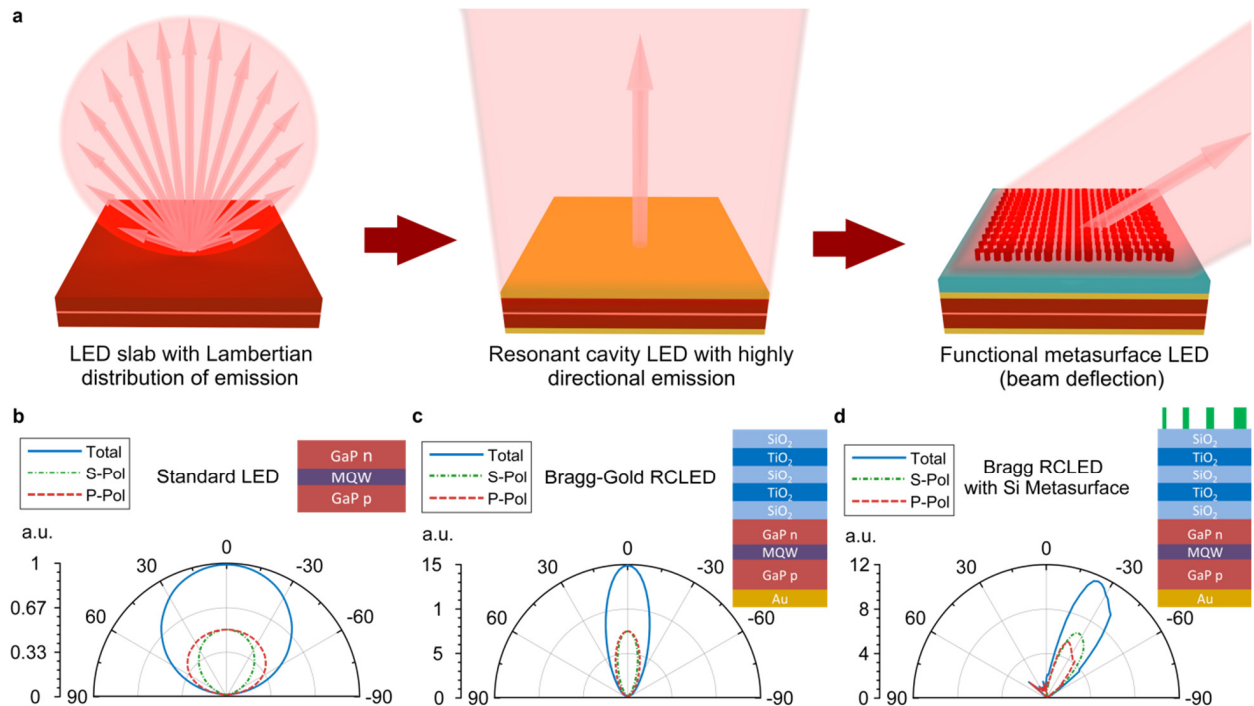

**Figure 1**. Working principle of the LED and metasurface integration and numerical simulations of emission diagrams of the LED in various configurations. a) the wide angle Lambertian distribution of the plain LED slab is transformed into highly directional emission using the resonant cavity. Next, metasurface is fabricated on top providing the desired functionality (beam deflection as an example). b, emission pattern of a semi-infinite GaP LED stack. Results show the characteristic Lambertian shape. c, emission pattern of a hybrid Bragg-gold RCLED. Results show directional emission towards the surface normal. d, emission pattern of the hybrid RCLED with the integrated, beam deflecting metasurface, as described in the main text (the polar plot shown corresponds to a cut of the full 3D pattern along the long axis of the metasurface supercell). Results show an emission lobe at 30° with respect to the normal, corresponding to the deflection angle of the metasurface when operating under plane wave excitation. All diagrams

are normalized to the maximum directional emission of the semi-infinite LED. The blue, solid line represents the unpolarized emission while the red and green dashed lines denote the p-polarized and s-polarized light components, respectively. The insets show the side view schematics of the different systems under consideration.

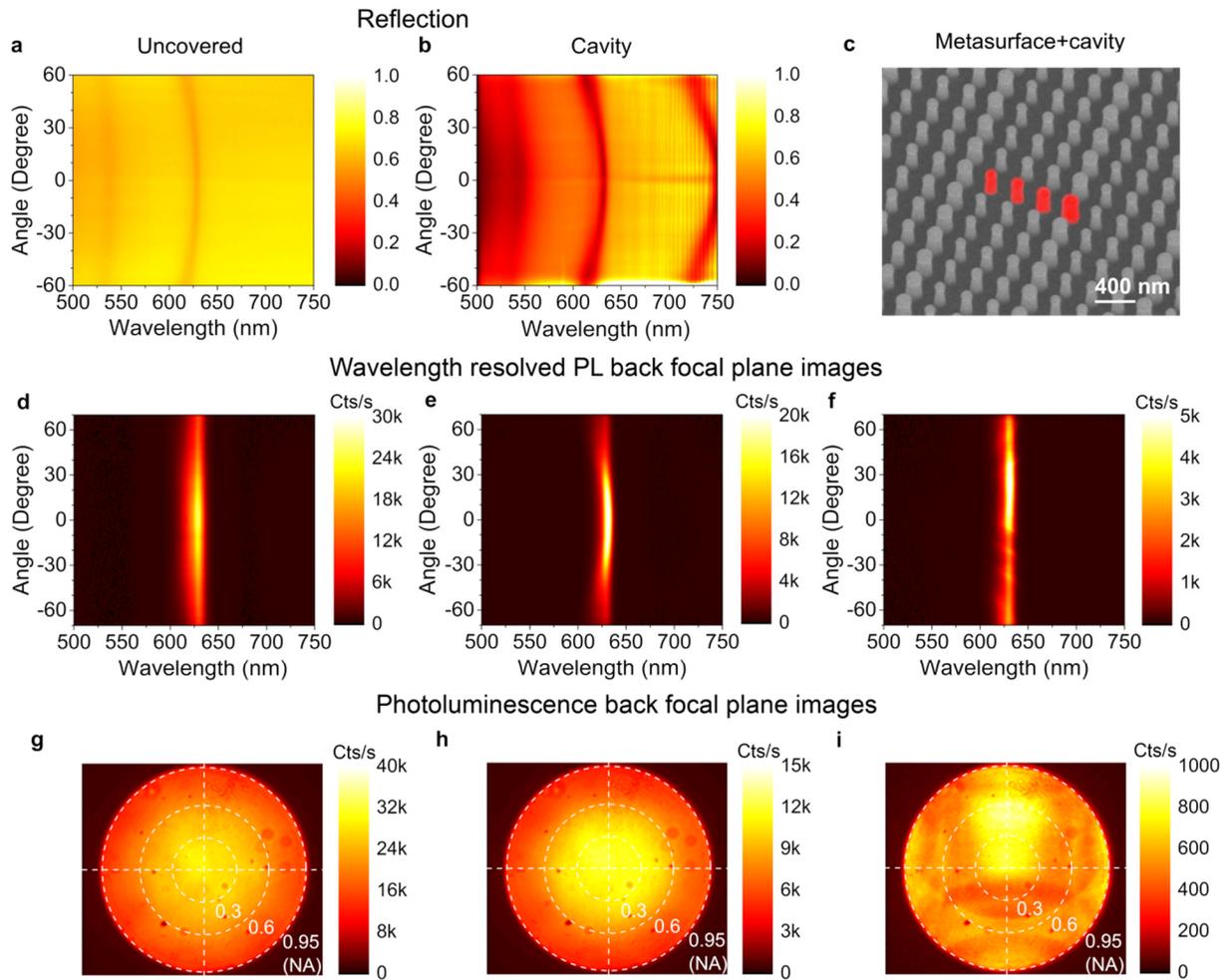

**Figure 2**. Experimental optical characterization of the uncovered LED and the resonant cavity LED without and with integrated beam deflecting metasurface. a-b, Angular-resolved reflection spectra of the (a) uncovered and (b) resonant cavity LED (RCLED). c, SEM image (taken at 30°) of the amorphous silicon-based, beam deflecting metasurface, fabricated on top of the RCLED. A supercell of the metasurface is highlighted in false red color. d-f, Angular resolved photoluminescence (PL) spectra of the uncovered LED (d) and the RCLED without (e) and with (f) the integrated beam deflecting metasurface on top. g-i, Back focal plane images of PL emission (including all wavelengths and polarizations) of the uncovered LED (g) and the RCLED without (h) and with (i) the integrated beam deflecting metasurface on top. PL was excited by a laser at the wavelength of 488 nm and collected using a microscope objective with NA=0.95.

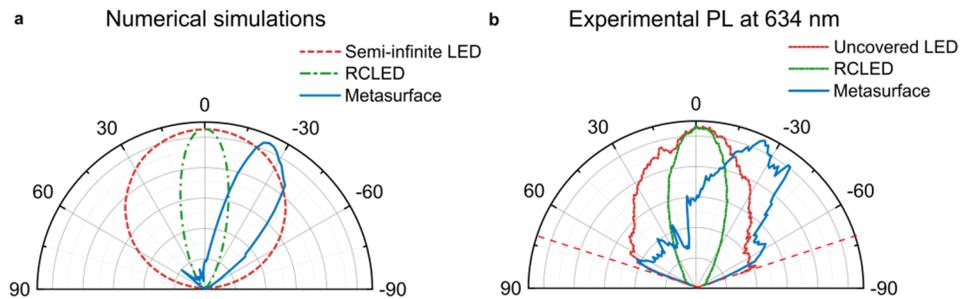

**Figure 3**. Comparison between the numerical simulations and experimentally measured emission directivity at a single wavelength. Simulated a) and measured b) emission patterns corresponding to the semi-infinite LED (red curve) and the RCLED without (green curve) and with (blue curve) the beam deflecting metasurface on the top. The experimental results correspond to the cross-section of the angular-resolved PL spectrum at 634 nm wavelength.

A resonant cavity increases the spacial coherence of the LED source and transforms a Lambertian profile into a more planar-like, suitable for most of the current state of the art metasurfaces excitation. This allows for direct and compact metasurface and LED device integration. The diversity of possible functionalities is demonstrated on the beam deflection and orbital angular momentum generating devices.

E. Khaidarov, Z. Liu[*], R. Paniagua-Domínguez, S. T. Ha, V. Valuckas, X. Liang, Y. Akimov, P. Bai, C. E. Png, H. V. Demir and A. I. Kuznetsov[*]

**Keywords**

dielectric metasurfaces, cavity, LED, vortex beam, directional emission

## Control of LED Emission with Functional Dielectric Metasurfaces

TOC Figure

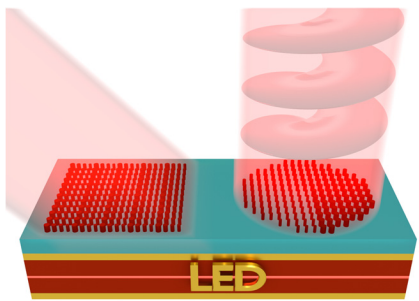

Supporting Information

## Control of LED Emission with Functional Dielectric Metasurfaces

*Egor Khaidarov[†], Zhengtong Liu[†*], Ramón Paniagua-Domínguez, Son Tung Ha, Vytautas Valuckas, Xinan Liang, Yuriy Akimov, Ping Bai, Ching Eng Png, Hilmi Volkan Demir and Arseniy I. Kuznetsov[*]*

*Numerical simulations of the metasurface and the gold cavity*

The amorphous Si metasurface was designed using the common phase mapping approach (see Ref. 29 in the main text) and simulated using a plane wave excitation. The resulting transmission, reflection and absorption spectra averaged over the two orthogonal incident polarizations are given in Figure S1a. The metasurface exhibits very low reflection in the wavelength range of interest and, hence, should have negligible effect on the properties of the collimating cavity. Almost all of the transmitted power is concentrated into the desired deflection direction (-1 diffraction order), corresponding to 94% of relative efficiency (normalized to the transmitted power) and 50% of absolute efficiency (normalized to the incident power) at the wavelength of interest of 620 nm.

Using the reciprocity principle we simulated the emission pattern of the plain LED with the Si metasurface on top (Figure S1b, size parameters of the metasurface are given in the main text). The shape of the emission is reminiscent of the Lambertian with slight deformations and clearly does not correspond to the designed beam deflection functionality. When we place the GaP LED inside a gold cavity (Figure S2a), it gives directionality similar to the hybrid gold-Bragg RCLED. Thickness of the cavity corresponds to the Fabry-Perot resonance (416 nm), top gold reflector is 50 nm thick, chosen as a tradeoff between high reflection and low absorption losses in metal.

The Si gradient metasurface placed on top of the gold RCLED deflects the emission at the designed 30° angle with respect to the normal as shown in Figure S2b. The emission diagram is similar to that of the hybrid Bragg-gold cavity case, but exhibits lower emission efficiency. The comparison of the emission efficiencies integrated over the top hemisphere for all proposed architectures is given in Figure S2c. The Bragg-gold RCLED has almost 3 times higher efficiency than the plain LED. The enhancement is reduced to 2 times when the metasurface is placed on top. Hybrid Bragg-gold RCLED with and without the metasurface have almost one order of magnitude higher efficiency than the pure gold one.

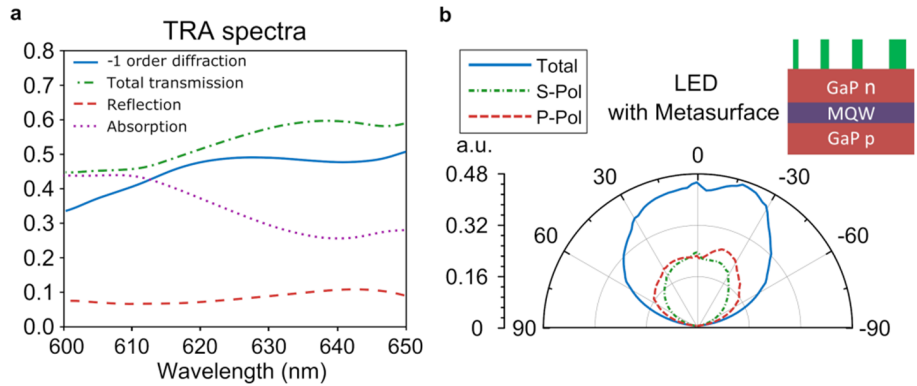

**Figure S1**. Numerical simulations of the metasurface performance and the emission diagram of the LED directly combined with the metasurface. a) Transmission, reflection and absorption (TRA) spectra of the Si metasurface excited with a normally incident plane wave; the metasurface is designed to deflect the incoming light at around 30° angle (into the -1 diffraction order) at 620 nm incident wavelength. a) Emission pattern of the plain semi-infinite LED with the Si metasurface directly on top. The shape of the emission is similar to the Lambertian one without the metasurface. The inset demonstrates the side view schematics of the structures, red and green lines in the emission patterns denote the p-polarized and s-polarized light components respectively, while the blue line denotes their sum.

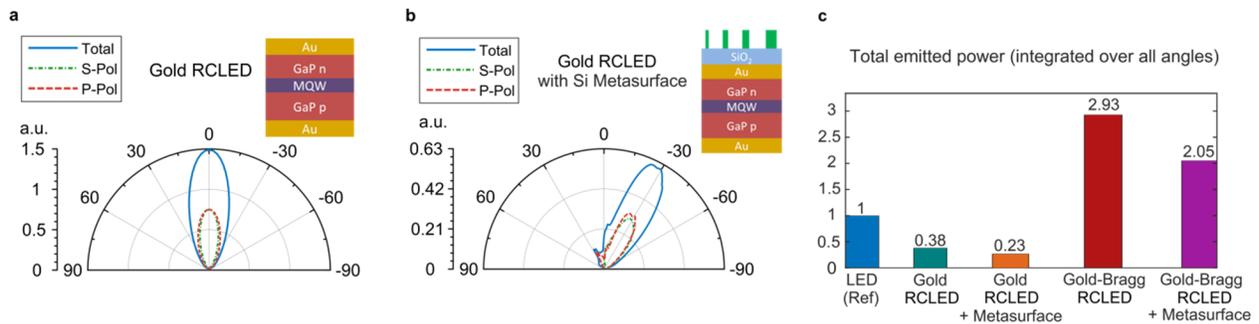

**Figure S2**. Numerical simulations of the emission diagrams of the RCLED combined with the metasurface and the total emission efficiency comparison. a) The emission diagram of the pure gold RCLED; the directionality is well pronounced. b) Directional emission of the gold cavity RCLED integrated with the Si metasurface (as described in the main text), the deflection angle of 30° corresponds to the theoretically designed angle for the plane wave excitation. The insets demonstrate the side view schematics of the structures, red and green lines in the emission patterns denote the p-polarized and s-polarized light components respectively, while the blue line denotes their sum. **c**, Total emission efficiency comparison for all proposed LED designs, calculated as an integral emitted power over the top hemisphere, normalized to that in the case of the plain semi-infinite LED stack.

*Vortex beam generation*

The vortex beam metasurface was fabricated using the same process as the beam deflecting metasurface. Generation of the orbital angular momentum +1 is achieved by continuous azimuthal variation of the cylinder radius, corresponding to 0-2π phase variation, around the geometrical center of the metasurface. SEM image of the central part of fabricated vortex beam generating metasurface is shown in Figure S3a. Photoluminescence (PL) was excited using a 590 nm laser from the metasurface side. The PL intensity has a distinctive donut shape profile shown in Figure S3b, characteristic for a vortex beam. Modulation of the emission intensity observed along the azimuthal direction can be caused by nonuniform absorption of the excitation beam in the metasurface for various pillar thicknesses, the maximum in intensity (the lowest absorption) corresponding to the thinner cylinders.

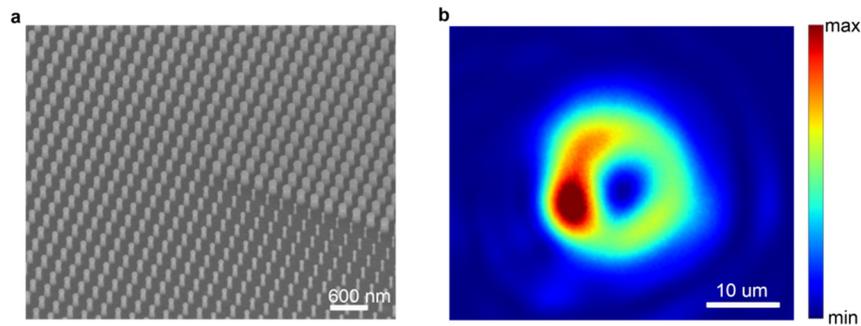

**Figure S3**. Experimental characterization of the GaP cavity LED with the vortex metasurface on top. a) SEM image of the central area of the vortex metasurface, taken at 30º tilt. b) Intensity profile of the PL from the vortex beam metasurface LED, excited at 590 nm wavelength.